\documentclass[aps,preprint,pre,showpacs,eqsecnum]{revtex4-1}
\usepackage{graphpap}
\usepackage{graphicx}
\usepackage{graphics}
\usepackage{color}
\usepackage{amssymb,amsmath}

\begin{document}

\title{Ultra-sensitive Hall sensors based on graphene\\ encapsulated in hexagonal boron nitride}

\author{Jan Dauber$^{1,2}$, Abhay A. Sagade$^{3}$, Martin Oellers$^{1}$, Kenji Watanabe$^{4}$, Takashi Taniguchi$^{4}$, Daniel Neumaier$^{3}$ and Christoph Stampfer$^{1,2}$}

\affiliation{
$^1$\,JARA-FIT and 2nd Institute of Physics, RWTH Aachen University, 52074 Aachen, Germany\\
$^2$\,Peter Gr\"unberg Institute (PGI-8/9), Forschungszentrum J\"ulich, 52425 J\"ulich, Germany\\
$^3$\,Advanced Microelectronic Center Aachen (AMICA) AMO GmbH, 52074 Aachen, Germany\\
$^4$\,National Institute for Materials Science, 1-1 Namiki, Tsukuba 305-0044, Japan
}
\date{ \today}

\begin{abstract}
The encapsulation of graphene in hexagonal boron nitride provides graphene on substrate with excellent material quality.  Here, we present the fabrication and characterization of Hall sensor elements based on graphene boron nitride heterostructures, where we gain from high mobility and low charge charier density at room temperature.  We show a detailed device characterization including Hall effect measurements under vacuum and ambient conditions. We achieve a current- and voltage-related sensitivity of up to $5700$~V/AT and $3$~V/VT, respectively, outpacing state-of-the-art silicon and III/V Hall sensor devices. Finally, we extract a magnetic resolution limited by low frequency electric noise of less than $50$~nT/$\sqrt{\textrm{Hz}}$ making our graphene sensors highly interesting for industrial applications.
\end{abstract}

 \pacs{???}  
 \maketitle
\newpage

Magnetic field sensors \cite{len90,rip10} are among the most widely used sensors with an annual production of around 5.9 billion units and prospects of further increase of ~50\% until 2020 \cite{mar14}. For example, in automotive and consumer electronics magnetic field sensor are heavily used for precise position detection, whereas these applications are dominated by Hall effect sensors. This fact is mainly due to their small size, high linearity and cost efficient production \cite{boe03}. The key performance indicators of Hall sensors are the magnetic resolution $B_{min}$ as well as the voltage-related ($S_V$) and current-related ($S_I$) sensitivities. These quantities depend crucially on the electronic properties of the active sensing material, such as the charge carrier mobility $\mu$ and the charge carrier density $n$ ( $S_V \propto \mu$ and $S_I \propto 1/n$ \cite{xu13}). Today, silicon based Hall sensors \cite{boe03,xu13,her93} are dominating most applications thanks to the well developed silicon CMOS technology, which enables efficient monolithic fabrication of the sensor element and the control electronics. However, for applications requiring higher sensitivity Hall sensors based on high mobility III/V semiconductors like GaAs or InAs are used \cite{har82,shi97,ber04,ban09} at the expense of higher fabrication
cost. Graphene \cite{gei07,nov12} is considered as the ideal material for ultra-sensitive Hall sensors because of its very high carrier mobility at room temperature and the ultra-thin body enabling very low carrier densities. These properties make graphene a material with the potential to outperform all currently existing Hall sensor technologies. Consequently, first graphene based Hall sensors have been demonstrated surpassing silicon based devices and approaching those based on III/V semiconductor materials in terms of sensitivity~\cite{hua14}. Importantly, the heterogeneous integration of graphene based devices on silicon CMOS substrates is feasible and has the potential of a cost efficient fabrication process \cite{hua14a}. 
\begin{figure}[tb]\centering 
\includegraphics[draft=false,keepaspectratio=true,clip,%
                   width=0.65\linewidth]%
                   {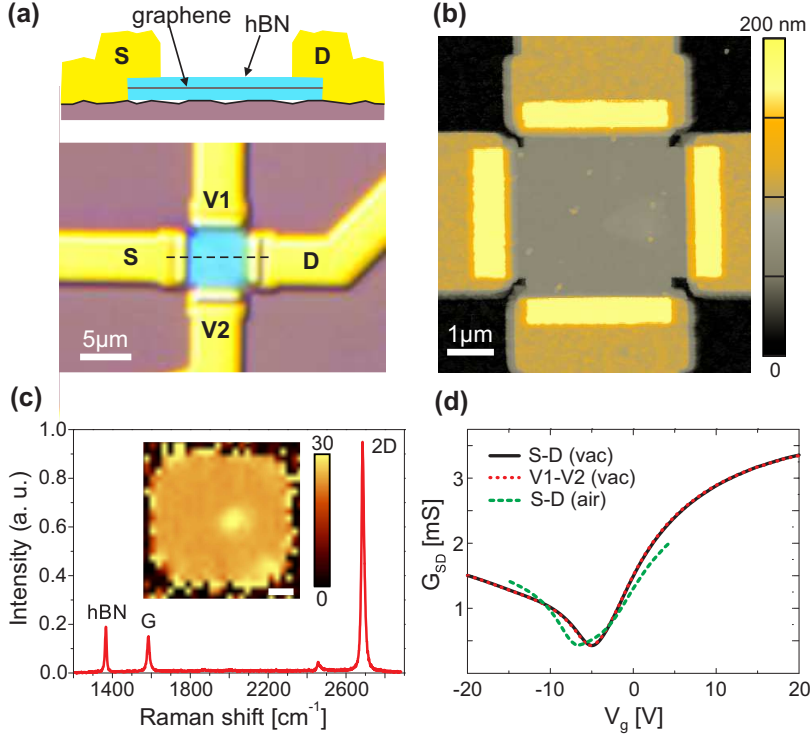}                   
\caption{
(a) Schematic illustration of a cross-section (top) and an optical image (bottom) of an etched and contacted graphene-hBN Hall sensor device (S1). Cross-section is shown along the dashed line in the optical image. (b) SFM image of the device S1 after fabrication. (c) Typical Raman spectrum obtained in the center of device S1.  The insert shows a Raman map of $\Gamma_{2D}$ in units of cm$^{-1}$.  The white scale bar is 1~$\mu$m. (d) Room temperature two-terminal back gate characteristics from device S1 under vacuum ($V_{b}=1m$~V) and ambient conditions ($V_{b}=5$~mV) for different contact configurations as indicated in panel (a).
} 
\label{figure1}
\end{figure}
With the encapsulation of graphene in hexagonal boron nitride (hBN) by van der Waals assembly \cite{wang13} graphene devices on substrate can be provided with very high mobility at room temperature and very low carrier density, both beneficial for the performance of Hall sensor elements. In this work we explore the performance limits of graphene based Hall sensors by utilizing high-mobility graphene-hBN heterostructures. In particular we show, that even with technically immature heterostructures we clearly surpass all existing state-of-the-art Hall sensors technologies.

The samples have been fabricated from mechanical exfoliated graphene, which is encapsulated in hBN using a stacking technique based on van der Waals adhesion \cite{wang13}. The hBN-graphene-hBN stacks are placed on highly p-doped Si substrates with a 285~nm thick SiO$_2$ top layer. Scanning force microscopy (SFM) and spatially resolved Raman spectroscopy \cite{gra07} are used to investigate the pristine heterostructures in order to identify homogeneous, residue-free graphene areas. In these regions an aluminum hard mask is patterned by electron beam lithography (EBL), metal evaporation (20~nm) and a lift-off process. The design is a symmetric cross with a width of 3~$\mu$m and a length of 5.2~$\mu$m. Subsequently, the uncovered hBN and graphene is removed by reactive ion etching with an SF$_6$/Ar plasma and the hard mask is stripped off by a wet chemical etching step. Contacts to the etched device have been made by a second EBL step, metal evaporation (5~nm Cr/95~nm Au) and a lift-off process.  An optical image and a schematic cross-section of a final device are depicted in Fig. 1(a). The SFM image (Fig. 1(b)) of the active device region shows a flat, homogeneous area with little contaminations. In Fig. 1(c) a single Raman spectrum as well as a Raman map of the full width at half maximum of the 2D peak $\Gamma_{2D}$ of the very same device are illustrated. The Raman spectrum not only proves that we are dealing with an isolated monolayer of graphene, but also provides insights on the material quality. In particular, the homogeneous $\Gamma_{2D}$ over the active device area (see insert in Fig. 1(c)) with a mean value of around $25$~cm$^{-1}$ is a good indication of a high material quality \cite{cou14,neu14}.

\begin{figure*}
\includegraphics[draft=false,keepaspectratio=true,clip,%
                   width=0.95\linewidth]
                   {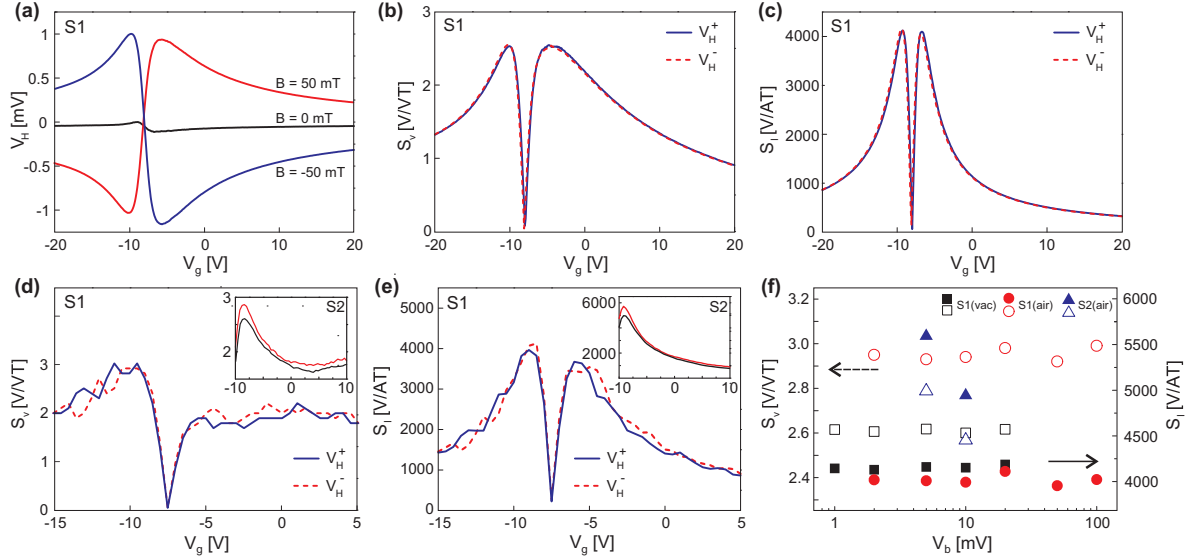}                   
\caption{
(a) V$_{H}$ as function of $V_g$ for different magnetic fields at $V_{b}$=10~mV under vacuum conditions. (b) and (c) Absolute value of $S_V$ and $S_I$ plotted against $V_g$ as derived from $V^{+}_{H}$ and $V^{-}_{H}$ at $V_{b}$=10~mV measured under vacuum. (d) and (e) Absolute value of $S_V$ and $S_I$ as function of $V_g$ derived from $V^{+}_{H}$ and $V^{-}_{H}$ at $V_{b}$=10~mV measured under ambient conditions. The inserts of these two panels show the corresponding data from a second device S2 at $V_{b}$=5~mV (red) and  $V_{b}$=10~mV (black) measured under ambient conditions. (f) Summary of maximum sensitivity S$_V$ (blank) and S$_I$ (filled) as function of $V_{b}$ for devices S1 and S2. 
}
\label{figure2}
\end{figure*}

The device (S1) shown in Fig. 1 is annealed in a tube oven under Ar/H$_2$ atmosphere at 275 $^\circ$C for 3h and first measurements are performed under low pressure Helium atmosphere ($\approx 5$~mbar) at room temperature, which is refereed in the following as under vacuum conditions (vac). In a next step, similar measurements have been performed under ambient conditions (air). As graphene is very sensitive to its environment and possible contaminations, we cover with these measurements both, laboratory as well as real life conditions. The two terminal back gate characteristics of the device under vacuum for the two different contact configurations exhibit a high similarity as depicted by the black and red (dotted) traces in Fig. 1(d), confirming the homogeneity as observed in the previous shown SFM and Raman data~\cite{SD}. The transport characteristics under ambient conditions (green dashed trace in Fig. 1(d)) reveal a shift of the charge neutrality point (CNP) to more negative back gate voltage and an asymmetric broadening of the minimum conductance dip around the CNP. These two effects can be attributed to contamination from the environment such as water, which may change the doping of the device.  Trace and retrace of the back gate characteristic show no significant shift or hysteresis as they overlap almost perfectly. For the determination of the magnetic response of the Hall sensor element a constant bias voltage $V_{b}$ is applied to contacts S and D.  The Hall voltage $V_{H}$ is measured between contacts V1 and V2 and the current $I_{SD}$ between contacts S and D  is recorded simultaneously. In Fig. 2(a) the Hall voltage $V_{H}$ for constant magnetic fields of $B=-50, 0\; \textrm{and}\; 50$~mT is plotted as function of back gate voltage $V_g$ under vacuum at $V_{b}$=10~mV. By subtracting the zero magnetic field values from the measured Hall voltages, $V^{\pm}_{H}=V_{H}(\pm50$~mT$)-V_{H}(0$~T$)$, we are able to suppress effects related to the geometry of the device and the measurement setup. Finally, we can use $V^{\pm}_{H}$ to extract the voltage-related $S_V$ and current-related $S_I$ sensitivities of our Hall sensor by utilizing the following expressions \cite{xu13},
\begin{align*}
 S_V= V_{b} \left| \frac{B}{V^{\pm}_{H}} \right| \hspace{1.5cm}  \textrm{and} \hspace{1.5cm}   S_I= I_{SD} \left| \frac{B}{V^{\pm}_{H}}\right|.
\end{align*}
In Figs. 2(b) and 2(c) we show the extracted sensitivities $S_V$ and $S_I$ as function of back gate voltage $V_g$ for a bias voltage of $V_b=10$~mV. The sensitivities depend crucially on charge carrier density with maximum values close the charge neutrality point. $S_V$ and $S_I$ show similar behavior for the hole and electron transport regime and are symmetric in magnetic field after applying $V^{\pm}_{H}$for the determination of the sensitivities. The maximum sensitivity is calculated as an average of the maximum values for holes and electrons derived from $V^{\pm}_{H}$ for $S_V$ and $S_I$, respectively. These measurements are repeated for bias voltages $V_b$ from 1 to 20~mV. Similar measurements are also preformed at a Hall probe station under ambient conditions with bias voltages $V_b$ ranging from $2$ to $100$~mV and magnetic fields of $B=-10, 0\; \textrm{and}\; 10$~mT. The data is processed in the very same manner as described above. In Figs. 2(d) and 2(e) the absolute values of $S_V$ and $S_I$ are plotted as function of back gate voltage $V_g$ for $V_{b}=10$~mV, respectively. $S_V$ and $S_I$ are again symmetric in magnetic field and no significant difference between trace and retrace is observed. However, $S_V$ exhibits now a hole-electron asymmetry around the CNP, where the absolute value of $S_V$ is increased for the hole regime and decreased for the electron regime with respect to measurements under vacuum.  In this case, only maximum values without any averaging are taken in account. In contrast, $S_I$ shows similar qualitative and quantitative behavior under ambient conditions as for measurements under vacuum. 
Additionally, we characterize a second device (S2) at a Hall probe station under ambient conditions. This device has not been annealed before measurements and it exhibits a stronger n-doping compared to device S1. Inserts of Figs. 2(d) and 2(e) show extracted sensitivities $S_V$ an $S_I$ from the second sample under ambient conditions. $S_V$ and $S_I$ exhibits similar or better performance, but both parameters decrease overall with bias voltage (see red and black traces in the insert). A summary of the maximum achieved sensitivities $S_V$ and $S_I$ for the different devices and measuring conditions as function of bias voltages are shown in Fig. 2(f). For device S1 the sensitivities are independent of applied bias voltage and $S_V$ increases by around $15\%$ from $S_V=2.6$~V/VT at vacuum to $S_V=3.0$~V/VT under ambient conditions,  whereas $S_I$ exhibits similar values of $S_I=4100$~V/AT for both measuring conditions. For devices S2 a sensitivity of up to $S_V=2.8$~V/VT and $S_I=5700$~V/AT is determined. This means that the relevant minimal accessible charge carrier density is independent on having the device under vacuum or ambient conditions, which is in contrast to the overall doping of the device (compare e.g. Figs. 2(b) and 2(e)). The increased $S_I$ for device S2 in contrast to device S1 indicates a smaller minimal accessible charge carrier density, which is also observed in Raman microscopy by a lower mean value of $\Gamma_{2D}$ \cite{raman}. The change in doping is most likely due to contaminations on the sample from the environment when measuring under ambient conditions. The increase and emerging asymmetry of $S_V$ when placing the device S1 from vacuum to ambient conditions may be the results of different doping level in the center area of the Hall cross and underneath the metal contacts. The significant difference of the bias dependency of both samples S1 and S2 can be most likely traced back to the missing annealing step for sample S2. Compared to annealed samples, this leaves an increased amount of resist residues on the surface of the top hBN-layer (confirmed by SFM images, but not shown). This may have influence on the device operation and stability. However, the details are far from being clear and further investigations on the influence of annealing remain important.

\begin{table}
	\centering
		\begin{tabular}{c c c c c c}\hline\hline
						&  $S_I$ & $S_V$ 	& $B_{min} \cdot w$ &	frequency &  conditions \\ 
				    &  [V/AT]  &  [V/VT]  & [pT/$\sqrt{Hz} \cdot mm$] & [kHz] & \\ \hline 
		Si \cite{pop03,ver09}   	
		&  100				 & 0.1 					& 1500 &	3	& NA	\\ 
		GaAs \cite{ver09} 
		&	 1100				 & NA						& 8000 &  3 & NA  \\ 
		InAsSb \cite{ban09}	
		&	 2750				 & NA 					& 50	 &  1 & NA  \\ 
		Graphene \cite{xu13,hua14}
				&  2093				 & 0.35			& 5000 &  3 & air  \\ 
	Gr-hBN (S1)		 &  4100 	& 2.6	  & 150  &  3 & vac \\
		Gr-hBN (S1)		 &  4000 	& 3.0	  & NA   & NA & air \\
		Gr-hBN (S2)		 &  5700 	& 2.8	  & NA   & NA & air \\	
		\hline\hline
		\end{tabular}
		\caption{Comparison of figures of merits of Hall sensors at room temperature reported in the literature and studied in this work ("Gr" stands for graphene).}
		\label{TABcomparsion}
\end{table}

\begin{figure}[t]\centering

\includegraphics[draft=false,keepaspectratio=true,clip,%
                   width=0.50\linewidth]%
                   {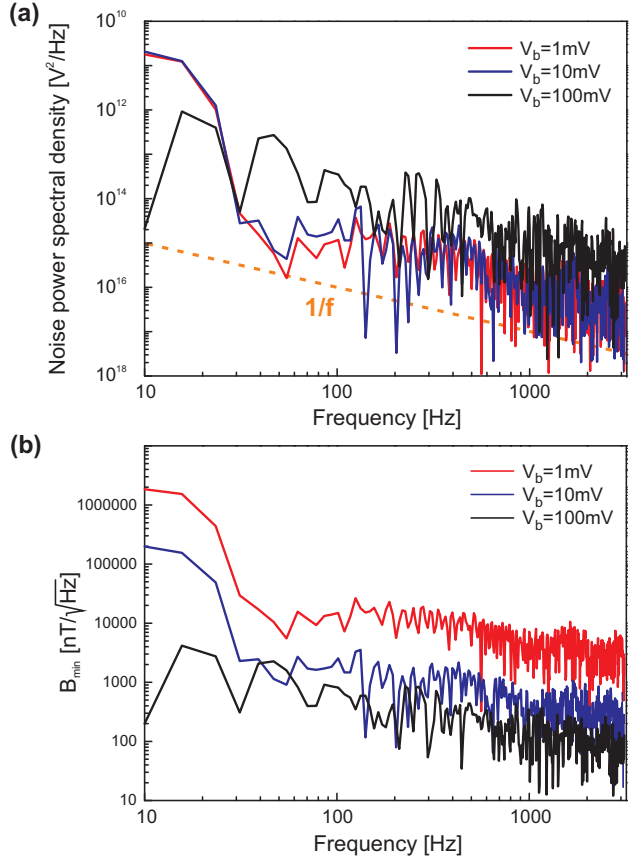}                   
\caption{
(a) Noise power spectral density $P_V$ as function of frequency. Orange dashed line indicates $1/f$ behavior. (b) Derived magnetic resolution $B_{min}$ of the Hall sensor element as function of frequency. 
}
\label{figure3}
\end{figure}
Besides the sensitivity, the magnetic resolution $B_{min}$ is a key performance parameter for applications, which is mainly limited by the charge noise of the device \cite{xu13a}. The noise power spectral density $P_V$ of the Hall voltage $V_{H}$ is measured (under vacuum) with a SR 770 FFT spectrum analyzer directly connected to contacts V1 and V2, while a constant bias voltage $V_{b}$ between contacts S and D and zero magnetic field $B$ are applied. In Fig. 3(a) we show $P_V$ as function of frequency for different bias voltages, where the expected $1/f$ dependence is well observed.  The magnetic resolution $B_{min}$ is derived from $P_V$ by using $B_{min}=\sqrt{(P_V)}/\left(S_V \cdot V_{b}\right)$ \cite{hua14} (see Fig. 3(b)). Since the sensitivity is constant for the measured bias voltage range, the magnetic resolution improves with increasing bias voltage and at 3~kHz and $V_{b}=100$~mV a magnetic resolution $B_{min}$ of $50$~nT$/\sqrt{Hz}$ is extracted. The non-linear increase of $P_V$ with bias voltage indicates, that the noise of the device is limited extrinsically by the measurement equipment and the above value can be taken as an upper estimate. For a better comparison with other Hall sensor elements the magnetic resolution is normalized by the contact width $w$, resulting in a normalized magnetic resolution $B_{min}\cdot w$ of $150$~pT/$\sqrt{Hz}\cdot $mm \cite{xu13}. 
In Tab. 1, we finally compare our findings with literature values for state-of-the-art Hall sensor elements made from silicon \cite{pop03,ver09}, III/V semiconductors \cite{ban09,ver09} and graphene \cite{xu13,hua14}. The table clearly shows that our graphene-hBN based Hall sensors are highly competitive with respect to all existing technologies. Most importantly, in terms of sensitivity $S_V$ and $S_I$ the present Hall sensor outperform silicon devices by more than one order of magnitude as well as III/V semiconductor and earlier graphene based devices by more than a factor of 2. Remarkably, the minimal magnetic resolution of our sensors is coming very close to the very best values achieved by InAsSb based sensors. All other material systems offer a significantly worse normalized magnetic field resolution.


In summary, we fabricated Hall sensor elements based on graphene-hBN heterostructures and characterize the magnetic response under vacuum and ambient conditions. We achieve a voltage-related sensitivity of up to $3.0$~V/VT and a current-related sensitivity of up to $5700$~V/AT surpassing not only silicon based Hall sensors, but also today's very best Hall sensors based on III/V semiconductors. These results unambiguously outlines the potential of graphene in commercial Hall sensor applications and clearly encourages efforts to improve the growth and transfer of CVD grown graphene in order to close the gap between exfoliated and CVD graphene. Developments towards a wafer scale fabrication of graphene-hBN heterostructures and their integration into CMOS technology is surely a very promising road to extend applications of future Hall sensor elements.

{Acknowledgments ---}
We thank U. Wichmann for help on electronics.  Support by the Helmholtz Nanoelectronic Facility
(HNF) at the Forschungszentrum J\"ulich, the EU Graphene Flagship project (contract no. NECT-ICT-
604391) and the ERC (GA-Nr. 280140) are gratefully acknowledged.


\begin{thebibliography}{99}
\bibitem{SD}
The measuring setup has a constantan wiring with a non-negligible resistance of about $R_W=300$~$\Omega$. The bias voltage $V_{b}$ denotes the applied voltage at the voltage source. The voltage drop across the Hall sensor element $V_{SD}$ is given by $V_{SD} = V_b - R_W*I_{SD}$.
\end{thebibliography}
\end{document}